\documentclass[runningheads]{svmult}

\usepackage{makeidx}   
\usepackage{graphicx}  
\usepackage{subeqnar}  
\usepackage{multicol}  
\usepackage{physprbb}  
\makeindex             

\def\lesssim{\mathrel{\hbox{\rlap{\hbox{\lower4pt\hbox{$\sim$}}}\hbox{$<$}}}}
\def\gtrsim{\mathrel{\hbox{\rlap{\hbox{\lower4pt\hbox{$\sim$}}}\hbox{$>$}}}}

\begin{document}
\title*{Red giant stars in NGC~5128}
\toctitle{Red giant stars in NGC~5128}
%
%
\titlerunning{Red giant stars in NGC~5128}
%
\author{Marina Rejkuba
}
%
%
%
\institute{European Southern Observatory, Karl-Schwarzschild-Str.~2, D-85748 Garching, 
	Germany}

\maketitle              

\begin{abstract}
I present a selection of results obtained from VLT FORS1 and ISAAC photometric
monitoring of late-type giants in NGC~5128 (=Centaurus~A). The combination of 
optical and near-IR photometry allows to probe the full metallicity range of the
stars on the upper red giant branch, thanks to combined low-metallicity sensitivity of the
optical and high-metallicity sensitivity of the near-IR bands. The 
metallicity distribution covers a wide range
with the mean value around $\mathrm{[M/H]} \sim -0.45$~dex. The near-IR monitoring
of the variable AGB stars allows to gain insights into the age 
distribution. The period 
distribution of these long period variables indicates only about 10\% 
contribution of the intermediate-age component ($\mathrm{age} \lesssim 5$~Gyr)
to the predominantly old stellar halo. Among the brightest, 
large amplitude and long period variables
only very few
have near-IR and optical colors consistent with carbon-rich giants. 
\end{abstract}

\section{Introduction and observations}

The red giant stars of NGC 5128 (=Centaurus A), 
the nearest, easily observable giant elliptical galaxy, were 
resolved first time by Soria et al.~(1996), who used WFPC2 camera on board HST to image
this galaxy's halo. Today, with the availability of 
8-10m class telescopes in excellent astronomical sites it is possible to obtain similar or even
better results with imaging from the ground. 

The data presented here have been obtained between April 1999 and July 2002 with FORS1
optical imager and spectrograph and ISAAC near-IR instrument
at UT1 Very Large Telescope at ESO Paranal Observatory. Two halo fields were observed 
once in $U$, $V$, $J_s$ and $H$ bands and monitored with 20--24 observations spread 
over 3 years in $K_s$. The observations, data reductions and photometric catalogues are 
presented by Rejkuba et al.~(2001, 2003a). Some results concerning the metallicity and
ages of red giant branch (RGB) and asymptotic giant branch (AGB) 
stars in the halo of NGC~5128 are shown. Here I plot color-magnitude and 
color-color diagrams for the north-eastern halo field (Field~1 in Rejkuba et al.~2001).
Very similar diagrams, and conclusions, are reached 
for the second, southern halo field, as well. For more detailed 
analysis, the interested reader is referred to Rejkuba et al.(2003a, 2003b) and 
Rejkuba (2004). 

\section{RGB and AGB in color-magnitude diagrams}

\begin{figure}[t]
\begin{center}
\includegraphics[angle=0,width=0.45\textwidth]{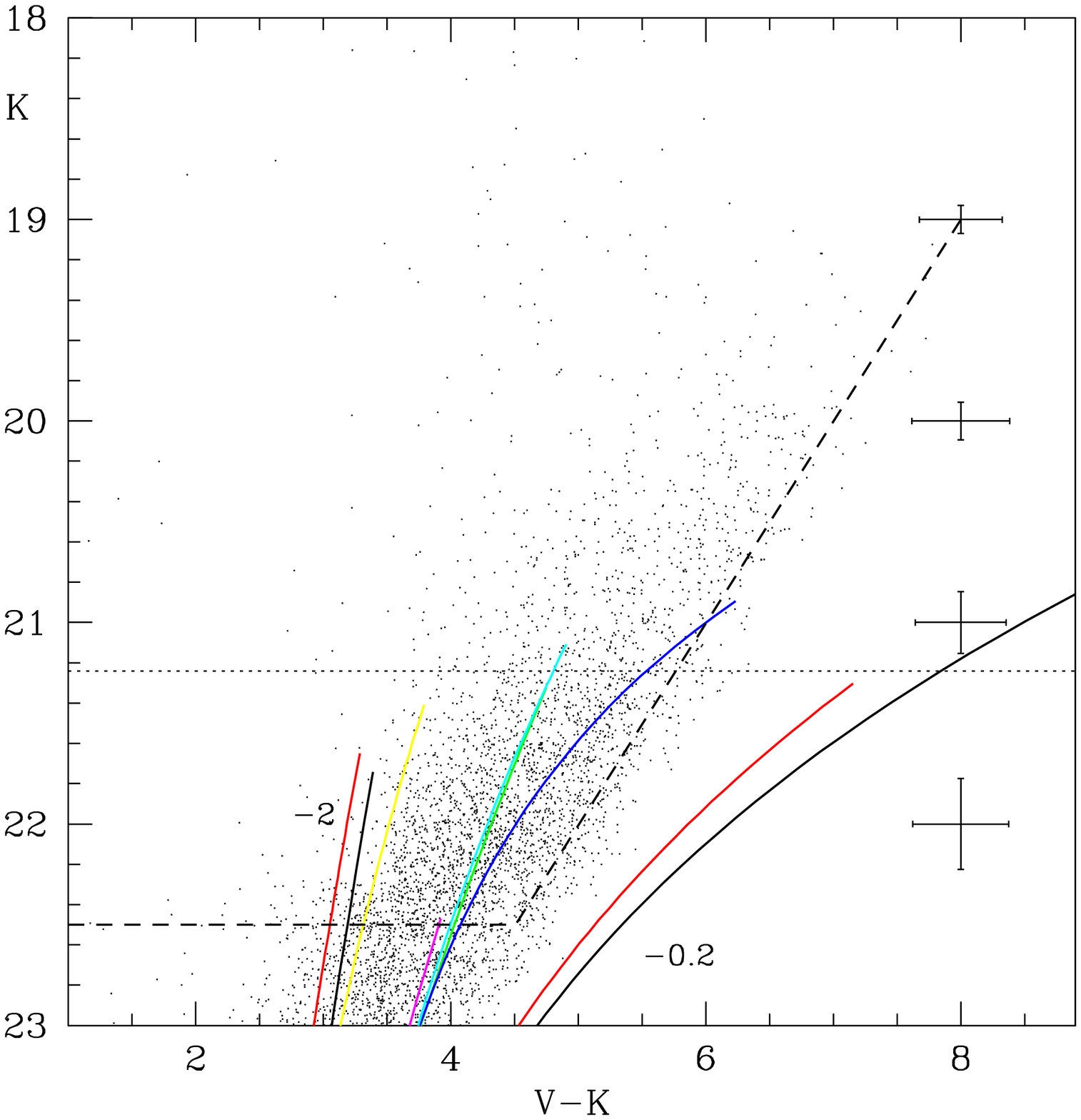}
\includegraphics[angle=0,width=0.45\textwidth]{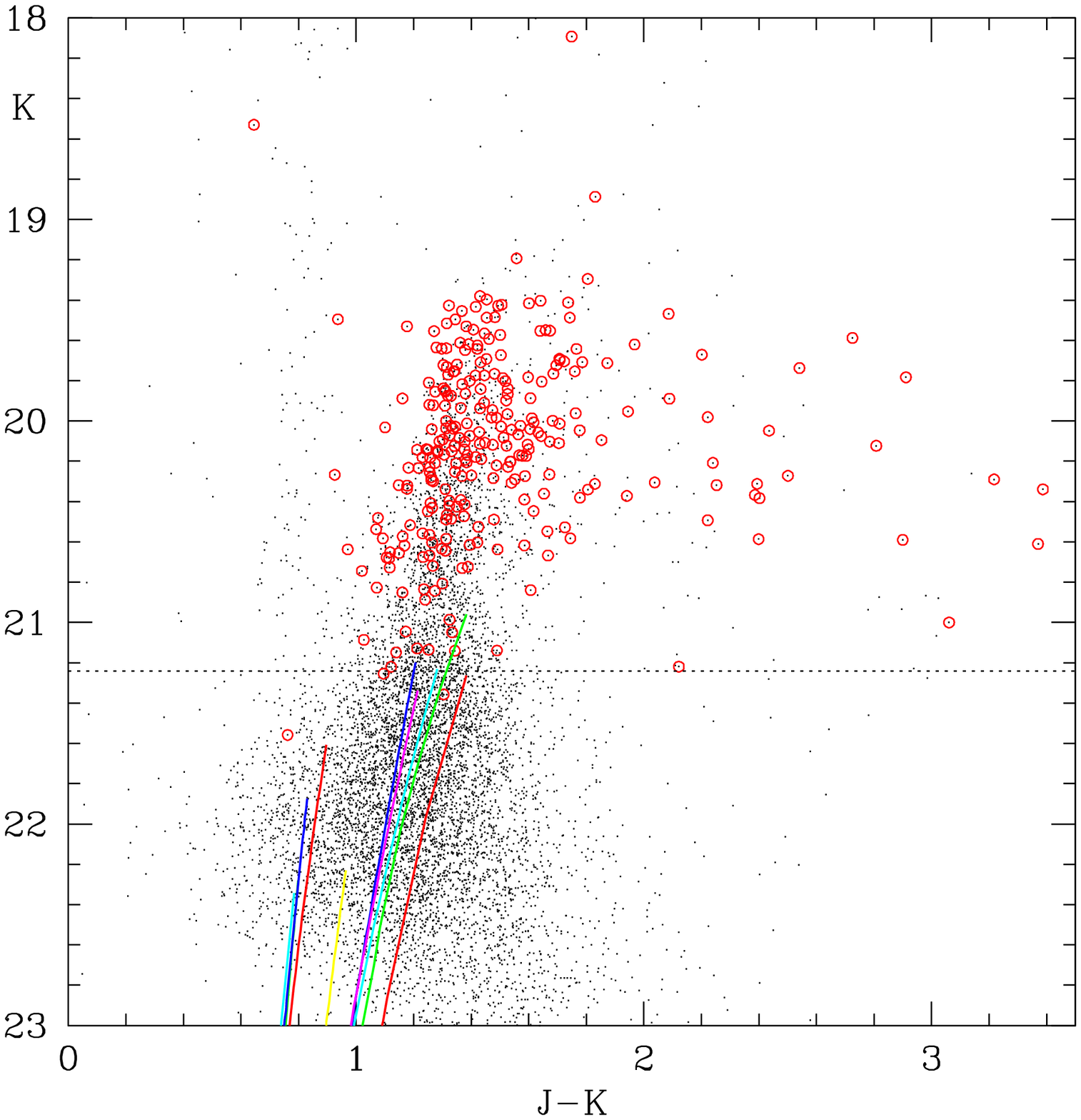}
\end{center}
\caption[]{{\bf Left:} Optical-near IR CMD of halo stars in 
NGC~5128. {\bf Right:} Near-IR CMD of the same field. \\
The horizontal line indicates the position of the RGB tip in both CMDs.
Overplotted are the fiducial RGBs for the following old Galactic 
globular clusters: NGC~6528, NGC~6553, M~69, 47~Tuc, M~107, M~4, M~55, M~30
and M~15, in the order of decreasing metallicity from $-0.2$ to $-1.9$~dex.
The dashed slanted line in the $VK$ CMD indicates 50\% 
completeness of the $V$-band photometry.
The large number of sources brighter than the RGB tip are AGB variables. Those variables
for which reliable periods could be obtained from our 3~yr-monitoring programme 
are marked with large circles.
}
\label{RGB}
\end{figure}

Combination of the optical and near-IR colors in a color-magnitude diagram (CMD) 
allows to probe old and intermediate-age stellar populations. Theoretically, more
than two thirds of the light in $K$-band is dominated by cool RGB and AGB stars. The
red dwarfs are too faint to be detected at the distance of NGC~5128, and thus the $VK$ and
$JK$ CMDs are entirely dominated by RGB and AGB stars (Fig.~1).

The spread in color of the RGB is larger than the photometric uncertainties, indicating 
the presence of a spread in metallicity and/or age. The age spread is possible, and indeed 
most likely (see below), but it cannot entirely account for the range of 
colors of the RGB stars, which are much more sensitive to metallicity changes. 
From a comparison with the fiducial RGBs 
of the old Galactic globular clusters (GGC) with a range of metallicities, and assuming 
old ages for the stars in the NGC~5128, the most metal-poor stars have 
metallicities as low as $-2$~dex, while the metal-rich end on the $VK$ CMD,  
set by the incompleteness of the $V$-band photometry (dashed slanted line), indicates 
approx.\ $-0.5$~dex. The most metal-rich giants, which are too faint in the
$V$-band due to huge bolometric corrections, emit most of their energy in the near-IR
and are thus easily observed in $JK$ CMD (Fig.~1, right). The most metal-rich GGC fiducials overplotted 
are for NGC~6553 and 6528, Bulge globulars, with 
$-0.3 \lesssim \mathrm{[M/H]} \lesssim -0.2$~dex. 
There is a small component of red giants in NGC~5128 with metallicities close to and 
slightly above solar. Their mean metallicity is $-0.45 \pm 0.05$~dex. 
For comparison, 
Walsh et al.~(1999) measured the mean oxygen-abundance of five planetary nebulae 
in NGC~5128 to be $\mathrm{[O/H]}=-0.5 \pm 0.3$~dex, 
in agreement with 
the average metallicity inferred from the RGB color. 

The dotted horizontal line in both CMDs in Fig.~1 is drawn at $K_s=21.24$~mag, 
the position of the tip of the RGB as measured from the ISAAC data. The
stars brighter than the RGB tip up to $\sim 2$~mag are in the AGB evolutionary phase.
There are additional few dozens of relatively bright stars in the $K$-band images with
no $J$, nor in $H$-band counterparts. Their stellar profiles and magnitudes
indicate that they might be highly dust obscured AGB stars.

\section{AGB variable stars}

\begin{figure}[t]
\begin{center}
\includegraphics[angle=270,width=.7\textwidth]{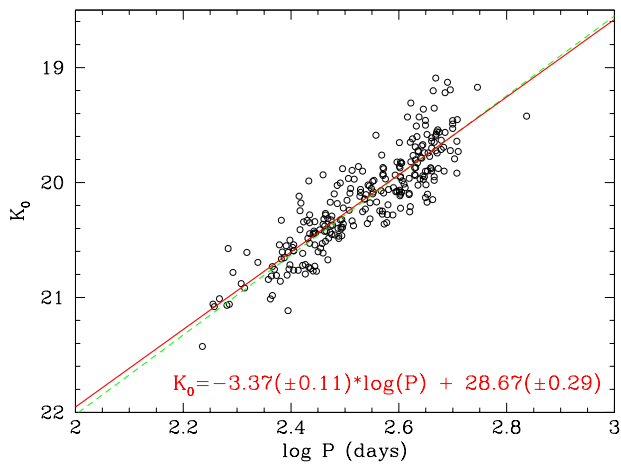}
\end{center}
\caption[]{Period-magnitude diagram for Mira LPVs in NGC~5128.
The solid line is a least-square fit to the data points (shown in the figure) 
and the dashed line is a 
fit with a fixed slope ($-3.47$) as measured in the LMC (Feast et al.\ 1989).
}
\end{figure}

Rejkuba et al.~(2003a) measured periods for more than 1000 long period variable (LPV)
NGC~5128 halo. The period-magnitude diagram for the LPVs with the most 
reliable periods is shown in Fig.~2. They form a sequence with the slope (indicated
in the figure) which is very close to that of the LMC Mira variable stars 
($-3.47$; Feast et al.\ 1989). Adopting the
zero point of the Mira period-luminosity relation of 0.98 and a distance 
modulus of the LMC of 18.50, 
I have obtained a distance modulus to NGC~5128 of $27.96 \pm 0.11$, in excellent
agreement with other literature values (see Rejkuba 2004 for more discussion).

\begin{figure}[t]
\begin{center}
\includegraphics[angle=270,width=.7\textwidth]{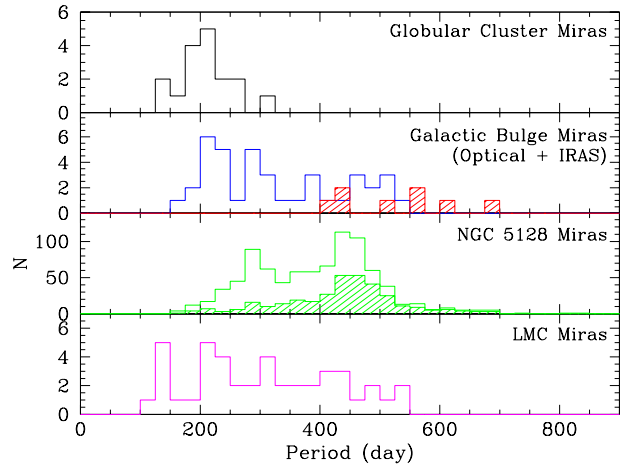}
\end{center}
\caption[]{Period distribution of LPVs in NGC~5128 is compared with: 
Galactic globular clusters ($\mathrm{age} \sim 12$ Gyr, $\mathrm{[M/H]}>-1$~dex),
Milky Way Bulge ($\mathrm{age} \sim 10$ Gyr, $\mathrm{[M/H]} \sim 0.0$~dex), and the
LMC (mostly $\mathrm{age}< 3$~Gyr,$\mathrm{[M/H]} \sim -0.7$~dex) Mira periods.
}
\end{figure}

Periods of Mira variables can be used to age date stellar population assuming that
their metallicities are known. Longer period Miras are expected to have higher 
mass progenitors, hence to be younger. However,
the more metal-rich, 
the older the star, assuming a constant mass. For example, according to models 
(e.g.\ Vassiliadis \& Wood 1993), a 1~M$\odot$ star will evolve
to a Mira variable with a period of $\sim 400$~days.
In a solar metallicity population the turn-off age of a 1~M$\odot$ star is 7.7~Gyr, 
while at $\mathrm{[M/H]}=-0.7$~dex it is 
$\sim 4.5$~Gyr.

A comparison of Mira period distributions
in NGC~5128 and 3 other systems with different mean ages and metallicities is 
shown in Fig.~3. The brightest Miras in NGC~5128 reach $M_K=-8.65$ and have periods in 
excess of 800 days. The large majority has similar periods to those of Galactic bulge
and old globular cluster Miras, but 10\% of them have periods in excess of 500 days
and are thus probably younger than $\sim 5$~Gyr, unless they all have extremely high metallicities. 

\section{Are there carbon stars?}

\begin{figure}[t]
\begin{center}
\includegraphics[angle=270,width=.7\textwidth]{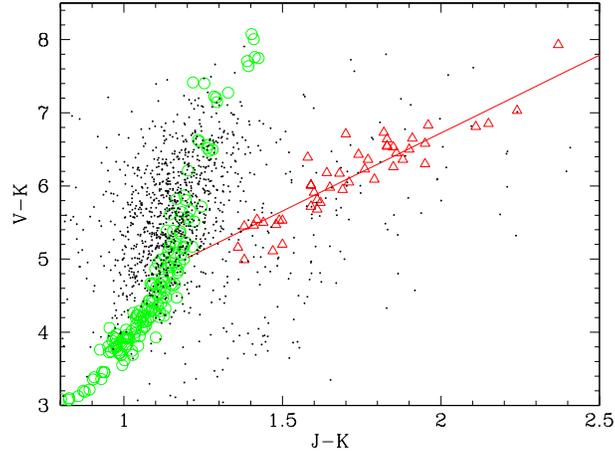}
\end{center}
\caption[]{Location of oxygen-rich (M-stars; large open circles;
Fluks et al.\ (1994) data) and carbon-rich (C-stars; large open triangles; Bergeat 
et al.\ (2001) data) giants from the Solar neighbourhood is compared with AGB stars
in NGC~5128 (small dots) in a optical near-IR color-color diagram.
The large majority of the AGB stars ($K_s<21.3$) in NGC~5128 are
oxygen-rich and only few have colors consistent with C-stars. Due to limits
of the $V$-band photometry, the reddest stars may be incomplete.
}
\end{figure}

Carbon stars are typically found among intermediate-age metal-poor populations. Their
presence is a definite proof of an intermediate-age component. If present in NGC~5128 
they may come from a recently accreted LMC-type or a small gas-rich spiral galaxy.
Carbon stars in the LMC have $1.4<(J-K_s)<2$, but the stars redder than $J-K_s>2$ can 
either be obscured oxygen or carbon-rich giants. $H-K$ vs. $J-H$ color-color diagram
has been a traditional tool for detecting these stars (see Rejkuba et al.~2003 for 
NGC~5128). In Fig.~4 I show a combined optical near-IR color-color diagram, $J-K$ vs.
$V-K$ for all the stars brighter than the RGB tip from the CMD in Fig.~1. The recent
models from Marigo (2002) fit the range of colors of late-type oxygen and carbon-rich
giants in the Solar neighbourhood in this diagram well. In Fig.~4 NGC~5128 AGB stars
are plotted with small filled symbols, Galactic M-stars are from Fluks et al.\ (1994; 
open large circles) and C-stars are from Bergeat et al. (2001; large open triangles).
Most of the stars in NGC~5128 are located along the oxygen-rich sequence, and only a small number of them are found along the location of carbon-rich giants. This is expected in 
a relatively high metallicity environment like that of NGC~5128 halo. However, as 
shown above there is also a metal-poor tail in this galaxy. Most of it appears to be 
older than $\sim 5$~Gyr, and only a very small component has intermediate-ages.

Much deeper photometry has been recently obtained with the ACS camera on board of HST. 
It reaches $V$-band magnitudes of $\sim 30$ and enables to detect additional 
age-sensitive features, like 
the AGB bump, red clump and the horizontal branch. The analysis of this
new observations will be used
to quantify the age distribution
of the NGC~5128 halo stars (Rejkuba et al., in preparation).

%


\begin{thebibliography}{7.}
\addcontentsline{toc}{section}{References}

Bergeat, J., Knapik, A, Rutily, B, 2001, A\&A, \textbf{369}, 178\\

Feast, M.~W., et al., 
1989, MNRAS, \textbf{241}, 375\\

Fluks, M.~A., et al., 1994, A\&AS, \textbf{105}, 311\\

Marigo, P., 2002, A\&A,  \textbf{387}, 507\\

Rejkuba, M., 2004, A\&A, \textbf{413}, 903\\

Rejkuba, M., Minniti, D., Silva, D.~R., Bedding, T.~R., 2001, A\&A, \textbf{379}, 781\\

Rejkuba, M., Minniti, D., Silva, D.~R., 2004, A\&A, \textbf{406}, 75\\

Rejkuba, M., Minniti, D., Silva, D.~R., Bedding, T.~R., 2004, A\&A, \textbf{411}, 351\\

Soria, R.~et al., 1996, ApJ,  \textbf{465}, 79\\

Vassiliadis, E., Wood, P.~R., 1993, ApJ, \textbf{413}, 641\\

Walsh, J.~R., Walton, N.~A., Jacoby, G.~H., Peletier, R.~F., 1999, A\&A, \textbf{346}, 754 


\end{thebibliography}
\end{document}